\documentclass[twocolumn,showpacs,showkeys,preprintnumbers]{revtex4}

\usepackage{dcolumn}
\usepackage{bm}
\input epsf
\begin{document}

\title{Inflation as a response to protect the Holographic Principle}
\author{V\'{\i}ctor H. C\'{a}rdenas}

\affiliation{Departamento de F\'{\i}sica y Astronom\'ia, Universidad
de Valpara\'iso, Av. Gran Bretana 1111, Valpara\'iso, Chile}

\begin{abstract}
A model where the inflationary phase emerges as a response to
protect the Fischler-Susskind holographic bound is described. A two
fluid model in a closed universe inflation picture is assumed, and a
discussion on conditions under which is possible to obtain an
additional exponential expansion phase as those currently observed
is given.
\end{abstract}

\pacs{98.80.Cq}

\keywords{inflation, physics of the early universe}

\maketitle

\section{Introduction}

From the release of the first studies by COBE \cite{cobe} on CMBR
fluctuations in 1994, cosmology started a period of precision
measurements that have re-shaped the subject. An important role in
the current understanding of the cosmological model is played by
Inflation \cite{inf,inf2,inf3}; the so called earlier stage after
the big bang where the universe expands at a quasi exponential rate.
Although inflationary models have proven to be a very successful
tool for cosmology, nobody knows why this period started. We only
know how this finished, evolving in a strong non-adiabatic and
out-of-equilibrium phase, called {\it reheating}. Because inflation
started at times as early as $10^{-35}\sec $, it is not difficult to
believe that inflation is a natural output coming from a Grand
Unified Theory (GUT) holding at Planck scale. So, we have no any
idea of how inflation started, because we cannot say anything about
such a theory.

However, 't Hooft \cite{thooft} proposed a crucial feature for that
theory: it has to be holographic. The idea of holography is the
following: if we want to reconcile quantum mechanics with gravity,
we have to assume that the observable degrees of freedom of the
universe are projections coming from a two-dimensional surface,
where the information is stored. During many years there has been
interest in the relation between holography and string theory
\cite{string,string2,string3,string4}.

The seminal work of Bekenstein \cite{bek,bek2} (see also
\cite{PG83}) on the universal upper bound on the entropy-energy
ratio for bounded systems suggested the idea of a holographic
universal (HP) principle that can be applied to cosmology. In this
context, Fischler and Susskind \cite{FS} proposed a holographic
principle and studied its consequences for the standard model of
cosmology.

From that time there has been a lot of work that attempts to refine
the original proposal of Fischler and Susskind (FS). This search has
tried to come up with a HP which does not conflict with inflation
and cosmology in general. In
\cite{holo,holo2,holo3,holo4,holo5,holo6} works on this line can be
found. Because the FS proposal was based on adiabatic evolution and
failed for closed Friedman-Robertson-Walker `cosmologies', the
interest was both extend the study to non-adiabatic evolution (as
reheating after inflation) and try to include the closed case. The
current understanding based on these studies implies that this
universal principle does not constraint inflation.

In Ref.\cite{EL,EL2} the authors propose to replace the HP by the
generalized second law of thermodynamics. The same idea was
developed in \cite{KL}, where the authors explicitly discussed the
problems of the FS proposal applied to inflationary models. More
recently, the subject has turned into its implication in dark energy
models. In \cite{miao} the author proposed a cosmological model
where dark energy can be accommodated within an holographic
principle, not only explaining the order of magnitude of dark energy
but also its equation of state \cite{DE}.

In general, holography has been used mainly in attempts to
adequately formulate an HP in cosmology. However, because the HP
must be considered a universal principle of a higher status than
inflation, in this work I look for a solution to the problem
considering a different point of view: an attempt to derive
inflation from the HP. This is so because we expect the HP, to come
from a (so far unknown) GUT theory, and then I will analyze if this
new point of view may constraint any inflationary model. From this
declaration of principles, it is clear that we do not find the
current beliefs very informative because inflation and the HP are
considered (almost) independent processes.

A first approach to this goal was presented in \cite{VHC02}, where I
proposed a model where inflation emerges from the saturation of the
HP in a closed FRW universe. The analysis was made taken care of
being thermodynamically consistent, as was stressed in \cite{KL} in
connection to the work of Rama \cite{rama}. Similar thoughts has
been suggested recently in \cite{dps07}.

In this work, I study a model where inflation can be interpreted as
a response to protect the violation of the HP {\it a la} FS, taking
care of the role of the curvature, and considering the non-adiabatic
process of reheating after inflation. I discuss the case of a closed
universe, which is apparently weakly favored by observations
\cite{closed}, with both radiation and a real scalar field $\varphi
$. At early times I found that an imminent violation of the HP
forces the system to saturate it through a period of exponential
growth of the scale factor. Also, I discuss the possibility to
explain additional exponential expansions phases like the one
observed now.

The results of a numerical integration of the evolution from
inflation to the present are shown in Figure 1. The key ingredient
to have an additional phase of accelerated expansion is reheating.
The extremely efficient transfer of energy from the scalar field to
radiation enables us to obtain a radiation-dominated phase with a
vanishing small (but not zero) relic of $\varphi $ fields. The
mechanisms through which this relic can be produced has been
discussed in \cite{relicfi,relicfi2,relicfi3,relicfi4,relicfi5}.
This relic does not confront observations because the universe
remains almost flat (until very small redshift) and during matter
domination, it is subdominant; however, it can dominate the matter
contributions for $z<0.5$. The holographic principle of Fishler and
Susskind is reviewed in section II. I present the model in section
III and develop the formulae to treat reheating in section IV. The
effect of protection of the HP is discussed in section V, and the
possibilities to have another phase of accelerated expansion is
discussed in section VI using the effect of plasma masses. The paper
ends with a discussion.

\section{The Fischler-Susskind Holographic Principle}

Let us assume a closed homogeneous isotropic universe with metric
\begin{equation}
ds^{2}=dt^{2}-a^{2}(t)(d\chi ^{2}+\sin ^{2}\chi d\Omega ^{2}),  \label{line}
\end{equation}
where $\chi $ is the azimuthal angle of $S_{3}$ and $\Omega $ is
the solid angle parametrizing the two-sphere at fixed $\chi $. The
particle horizon is
\begin{equation}
\chi _{H}=\int_{t_{i}}^{t}\frac{dt^{\prime }}{a(t^{\prime })},  \label{horiz}
\end{equation}
where $t_{i}$ is a reference initial time. Because integral
(\ref{horiz}) may diverge at small $t$, a natural choice is to
take $t_{i}=t_{pl}=1$ \cite{KL}. The angle $\chi _{H}$ determines
the coordinate size of the horizon, which defines a bounded area
and volume. Because the entropy density $\sigma \equiv (\rho
+p)/T$ is constant, the entropy area ratio gives
\begin{equation}
\frac{S}{A}=\sigma \frac{2\chi _{H}-\sin 2\chi _{H}}{4a^{2}(\chi _{H})\sin
^{2}(\chi _{H})}.  \label{s/a}
\end{equation}
As the universe evolves, ratio (\ref{s/a}) increases and the
system reaches a stage of saturation and later, a violation of
holographic principle \cite{FS}. For example, for a universe
filled with non-relativistic matter, $a=a_{\max }\sin ^{2}(\chi
_{H}/2)$ so for its maximal expansion $\chi _{H}=\pi $ ratio
(\ref{s/a}) becomes violated.

\section{The model}

I model the universe as filled with both, a single scalar field -
inflaton $\varphi $ - and a fluid of relativistic particles with
energy density $\rho _{m}$. Assuming a homogeneous field with a
slowly-time-dependent equation of state $p_{\varphi }=w(t)\rho
_{\varphi }$, where
\begin{equation}
w(t)=\frac{\dot{\varphi}^{2}-2V(\varphi )}{\dot{\varphi}^{2}+2V(\varphi )},
\label{ratio}
\end{equation}
we can write the Friedman equation as
\begin{equation}
H^{2}=H_{0}^{2}\left[ \Omega _{\varphi }\left( \frac{a_{0}}{a}\right)
^{3(1+w)}+\Omega _{m}\left( \frac{a_{0}}{a}\right) ^{4}-\Omega _{k}\left(
\frac{a_{0}}{a}\right) ^{2}\right] ,  \label{hubble}
\end{equation}
where $a_{0}$ and $H_{0}$ are the scale factor and the Hubble parameter in
an arbitrary reference time $t_{0}$ (often taken to be the present time).
Also, we have defined the following parameters: the density parameter $%
\Omega _{i}=\rho _{0i}/\rho _{c}$ for the $i$ component with
initial energy density $\rho _{0i}$; the critical density $\rho
_{c}=3H_{0}^{2}M_{p}^{2}/8\pi $, the equivalent energy density due
to
curvature $\rho _{k}=-3M_{p}^{2}/8\pi a^{2}$; and the Planck mass $%
M_{p}=1.2\times 10^{19}$GeV.

Because I am considering a pressure density ratio in the range $-1<w<-1/3$%
, this implies that at some time $t^{\ast }$ the first term in the
square brackets of (\ref{hubble}) will dominate over the other
contributions,
starting a period of `inflationary expansion' (if we also have $\ddot{a}>0$%
). This exponential expansion leads to a process of `flattening' of
the universe due to the screening of the curvature term. The authors
of \cite{KT2} used the latter effect to obtain an apparent spatially
flat FRW universe by using a closed one.

If we extend the evolution to the present (i.e., at $t_{0}$), the
evolution of the system leads to a total density parameter
\[
\Omega =\sum_{i}\Omega _{i}=\Omega _{\varphi }+\Omega _{m}=1+\Omega
_{k}\simeq 1,
\]
because $\Omega _{k}=\rho _{0k}/\rho _{c}=3M_{p}^{2}/8\pi
a_{0}^{2}H_{0}^{2}\ll 1$ when we evaluate it today. Globally, the
effect agrees with the results of \cite{rama}, which to solve the
violation problem of the HP bound for a closed universe,
introduced an `exotic fluid' - as the inflaton here - as an extra
matter component. However, the presence of this component leads to
a period of non-adiabatic evolution \cite{KL}, so the equations of
motion must be corrected to consider entropy production during the
reheating process.

The simplest way to do that is to consider - as a first
approximation - the perturbative regime of reheating which was
described by adding a friction term to the inflaton equation of
motion \cite{REHEAT,REHEAT2,REHEAT3,REHEAT4}. In this paper, I have
taken into account the most efficient energy transfer possible,
based on the main results of the modern reheating theory
\cite{MthR,MthR2,MthR3,MthR4}.

\section{Particle and/or entropy production}

In Ref.\cite{KL} Kaloper and Linde concluded that the proposal of
Fischler and Susskind does not confront inflation, because it
eliminates the entropy produced inside the light cone during
reheating. This means that their HP is valid during adiabatic
expansion. To extend the analysis through the particle creation
process, we have to learn how to compute ratio (\ref {s/a}) in a
non-adiabatic stage, i.e., during reheating.

I consider the simplest model of reheating, which is based on the
Born approximation \cite{REHEAT}. Here, the inflaton evolves
according to
\begin{equation}
\ddot{\varphi}+3H\dot{\varphi}+V^{\prime }(\varphi )=-\Gamma
\dot{\varphi}, \label{ieom}
\end{equation}
where $\Gamma $ is the rate of particle production, and the
evolution of the relativistic particles created is described by
\begin{equation}
\dot{\rho}_{m}+4H\rho _{m}=\Gamma \rho _{\varphi }.  \label{reom}
\end{equation}
Equations (\ref{ieom}) and (\ref{reom}) explicitly show the
non-adiabatic nature of the process. If, as usual, we define the
energy density and pressure by
\begin{equation}
\rho =\frac{1}{2}\dot{\varphi}^{2}+V(\varphi ),\ p=\frac{1}{2}\dot{\varphi}%
^{2}-V(\varphi ),  \label{eq2}
\end{equation}
we can write the equation of state during the rapid oscillations
phase of $\varphi $, i.e., $V^{\prime \prime }\gg H^{2}$ as a
temporal average during an oscillation by $\langle (\rho + p)/
\rho \rangle = w=(n-2)/(n+2)$ where $V(\varphi )=\varphi ^{n}$, so
\begin{equation}
\dot{\rho}_{\varphi }+3H\rho _{\varphi }(1+w)=-\Gamma \rho _{\varphi }.
\label{eq3}
\end{equation}
In the special case $w=0$ (a quadratic potential), this equation
describes the decay of a massive particle. A solution of this
equation is
\begin{equation}
\rho _{\varphi }=M^{4}\left( \frac{a_{i}}{a}\right) ^{3}\exp \left[ -\Gamma
\left( t-t_{i}\right) \right] ,  \label{eq4}
\end{equation}
where subscript $i$ indicates the epoch when the coherent
oscillations around the minimum of the potential $V(\varphi )$
begins, and $M^{4}$ is the vacuum energy at that time.

If the produced particles are thermalized, we can use the
expression for the entropy of radiation
\begin{equation}
S=\frac{2\pi ^{2}}{45}gT^{3}a^{3},  \label{eq6}
\end{equation}
which combined with the equality $\rho _{m}=\pi ^{2}gT^{4}/30$,
valid for relativistic particles and Eq.(\ref{eq6}), enables us to
write a relation between the energy density and the entropy
\begin{equation}
\rho _{m}=\frac{3}{4}\left( \frac{45}{2\pi ^{2}g}\right) ^{1/3}S^{4/3}a^{-4}.
\label{eq9}
\end{equation}
For $t<\Gamma ^{-1}$, the universe is dominated by $\varphi $
particles and, according to Eq.(\ref{hubble}) and (\ref{eq4}), it
evolves as a matter-dominated universe $a(t)\sim t^{2/3}$. An
approximated solution of (\ref {reom}) is
\[
\rho _{m}\simeq \frac{1}{10}M_{p}\Gamma M^{2}\left( \frac{a}{a_{i}}\right)
^{-4}\left[ \left( \frac{a}{a_{i}}\right) ^{5/2}-1\right] ,
\]
which implies that, initially $\rho _{m}$ increases from \ $0$ to $%
M_{p}^{2}\Gamma M^{2}$, and after that it decreases as $a^{-3/2}$
(see Figure 1). From (\ref{eq9}) we find that the entropy grows as
$S\propto a^{15/8}$. This means that $\sigma $ (which appears in
front of (\ref{s/a})) is no longer constant, and has to be
replaced by a function varying as $\sim $ $a^{15/8}$\cite
{turner85}.

\begin{figure}[t]
\centering \leavevmode\epsfysize=9cm \epsfbox{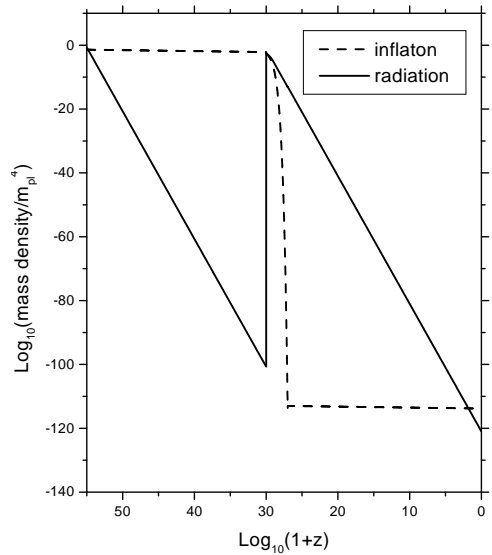}\\
\caption[fig1]{\label{fig1} Evolution in the model. Inflation
starts at $z=55$ and finishes at $z=30$. Note that at $z\sim 1$,
the inflaton field starts to dominate the matter content leading
to a new exponential expansion phase.}
\end{figure}

\section{Protecting the HP with Inflation}

Let us assume that the universe starts in a radiation dominated
era $\rho _{m}\sim \rho _{k}\gg \rho _{\varphi }$ at a time near
the beginning of inflation, say $10^{-35}\sec $. As was discussed
in Section II, such a universe will evolve towards an imminent
violation of the HP. In fact, during this phase Eq.(\ref{hubble})
can be written as
\begin{equation}
H^{2}=\frac{8\pi }{3M_{p}^{2}}\rho _{0m}\left( \frac{a_{0}}{a}\right) ^{4}-%
\frac{1}{a^{2}},  \label{initH}
\end{equation}
from which we obtain the solution
\begin{equation}
a(\chi _{H})=A\sin (\chi _{H}),  \label{sol1}
\end{equation}
where $A\equiv \sqrt{8\pi \rho _{0m}a_{0}^{4}/3M_{p}^{2}\text{ }}$. By
inserting (\ref{sol1}) in (\ref{s/a}), we find that the HP is violated as $%
\chi _{H}\rightarrow \pi $. Note that in this form, the bound
(\ref{s/a}) is violated also as $\chi _{H}\rightarrow 0$. This
particular problem is solved by the arguments displayed under
Eq.(\ref{horiz}). Because $a(\chi _{H}=0)=a_{p}\neq 0$, then it is
possible to write the solution (\ref{sol1}) as $a(\chi
_{H})=a_{p}+A\sin (\chi _{H})$, solving the HP bound (\ref{s/a})
as $\chi _{H}\rightarrow 0$. However, the violation in the future
($\chi _{H}\rightarrow \pi $) remains.

Now, let us follow the evolution of both (\ref{s/a}) and
(\ref{hubble}) during the transit from radiation domination to the
inflationary phase. If we assume at Planck time $t_{pl}$ that
$\rho _{m}\sim \rho _{k}$, then we can expect after certain time
say $t\simeq 10^{-35}\sec $, to enter an inflationary phase. In
fact, because $\rho _{\varphi }\sim a^{-3(1+w)}$ with $-1<w<-1/3$
decays more slowly than $\rho _{k}\sim a^{-2}$and $\rho _{m}\sim
a^{-4}$, the former dominates. If we consider the beginning of
inflation when $\rho _{\varphi }$ becomes comparable to $\rho
_{k}$, this implies that at the time $t\sim 10^{-37}\sec $ we have
\[
\rho _{\varphi }\simeq 10^{-2}(10^{-1})\rho _{k},
\]
for $w=-1$ ($-2/3$), respectively. After $\rho _{\varphi }$
becomes greater than $\rho _{k}$, this component dominates the
matter content in the universe and makes it inflationary in the
sense that $\rho _{\varphi }\gg \rho _{k},\rho _{m}$. Here the
Hubble parameter $H$ (\ref{hubble}) becomes nearly constant, and
the coordinate size of the horizon (\ref{horiz}) behaves like
\begin{equation}
\chi _{H}=\int_{t_{p}}^{t}\frac{dt^{\prime }}{a(t^{\prime })}\simeq
H^{-1}(e^{-Ht_{p}}-e^{-Ht}),  \label{hosat}
\end{equation}
reaching an asymptotic constant value proportional to $H^{-1}$.
Because the scale factor grows nearly exponentially, the HP bound
(\ref{s/a}) starts to decrease, avoiding the violation in the
future. A numerical integration of the system is shown in Fig. 2.
Specifically we plot the entropy bound (\ref{s/a}) together with the
energy density of both radiation and the inflaton field. As is shown
in the figure, before the onset of inflation, the entropy bound is
not saturated however this evolves towards saturation. Once the
violation of the bound is imminent, inflation starts modifying the
evolution avoiding the violation. In this way, the appearance of
inflation saves the violation of the HP in the future. For example,
in the $w=-1$
case, the scale factor can be written as $a(\chi _{H})\simeq \sqrt{3/8\pi }%
\left[ \sin (\pi /2-\chi _{H})\right] ^{-2}$, which behaves much
like an exponential growth. The early radiation dominated phase is
not necessary at all to demonstrate the role of the HP in
inflating the universe. In fact, we can start the universe with a
bound $S/A$ saturated at the Planck era; this time with $\rho
_{\varphi }\geq \rho _{k},\rho _{m}$.

\begin{figure}[t]
\centering \leavevmode\epsfysize=9cm \epsfbox{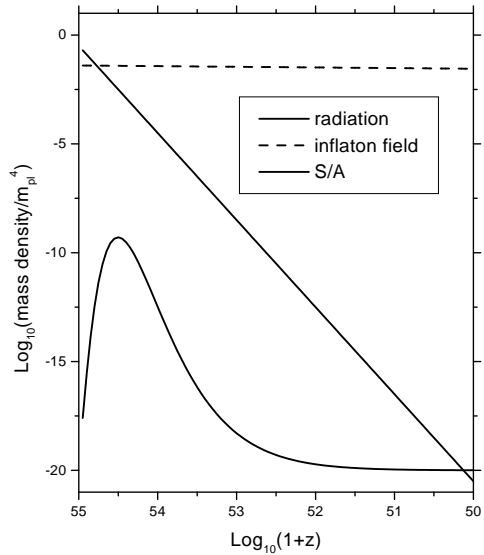}\\
\caption[graph1]{\label{graph1} This plot shows a detailed
evolution of the energy densities and the holographic bound at
early times. This shows how inflation appears to protect the
violation of the HP.}
\end{figure}

To solve the cosmological problems, inflation must last around
$60$ e-folds. This means $a_{end}=e^{60}a_{init}$. This growth
makes the curvature term irrelevant for the subsequent analysis,
although it was fundamental at
the beginning. For example, if we assume that inflation begins when $%
H^{2}\sim a^{-2}$ in Eq.(\ref{hubble}), then after the $60$ $e$-folding we
have in Eq.(\ref{hubble}) $a^{-2}\simeq e^{-120}H^{2}\ll H^{2}$, making the
universe very flat.

After this 60 e-folding of inflation, particle creation starts.
During this phase the formulae derived in section IV are valid,
and the expression for the HP bound has to be replaced by
\[
\frac{S}{A}=\sigma \frac{\chi _{H}}{a^{2}}\propto \frac{H^{-1}}{a^{1/8}},
\]
which decreases more slowly than in the previous phase. Because
the curvature term has fallen more than $50$ orders of magnitude
from the beginning of inflation, it seems difficult to repeat a
similar argument to explain the current accelerated expansion.
This possibility is discussed in the next section.

\section{Possible extra inflationary stages}

After the 60 $e$-folding of inflation, bound (\ref{s/a}) has fallen
to a negligible small value making impossible the existence of HP
saturation for another period. However, we have seen that after
reheating the universe becomes radiation-dominated, and we can
consider a small relic of the inflation field, which enables us to
have another accelerated period. This possibility is open if the
inflaton energy density does not fall more than 120 orders of
magnitude during reheating.

To get a viable observational model we have to consider the process
of reheating once the parametric resonant phase
\cite{MthR,MthR2,MthR3,MthR4} has finished. In this new context, the
study of reheating is clearly different from the standard scenario;
now the universe not only has the contribution of the coherent
oscillations of the inflaton field but also the particles created
during preheating. Therefore, our problem is to find a mechanism
that not only allows to avoid that the field inflaton decays
completely, but also take into account the presence of particles
created in the previous phase.

In Ref.\cite{relicfi3} I proposed a way to obtain this incomplete
decay of the inflaton field. In that scenario we assumed that during
the process of preheating the inflaton decay products scattered and
thermalize to form a thermal background \cite{KNR}. This thermalized
particle species acquires a plasma mass $m_p (T)$ of the order of
$\nu T$ where $\nu $ is the typical coupling governing the particle
interaction. The presence of thermal masses imply that the inflaton
zero mode cannot decay into light states if its mass $m$ is smaller
than about $\nu T$. If we expressed the inflaton zero-temperature
decay width as $\Gamma_{\phi}=\alpha_{\phi} m$, at finite
temperature it becomes
\begin{equation}\label{gamat}
\Gamma_{\phi}(T)= \alpha_{\phi}m \sqrt{1-4\frac{\nu^2 T^2}{m^2}}.
\end{equation}
The system of equations to be solved is then
\begin{equation}\label{syseq}
\begin{array}{c}
          \dot{\rho_{\phi}}+ (3H + \Gamma_{\phi} )\rho_{\phi}=0, \\
          \dot{\rho_{R}}+ (4H\rho_{R} - \Gamma_{\phi} )\rho_{\phi}=0.
\end{array}
\end{equation}
When the plasma mass $m_p (T)=\nu T$ becomes comparable to the
inflaton mass $m$, the temperature reaches the value $T \simeq
m/2\nu$, remaining constant for a while, indicating that particle
creation stopped; $\Gamma_{\phi}=0$. At this time $\rho_R$ stays
constant and $\rho_{\phi}$ decays as $a^{-3}$.

Based on observations, we have some good evidence of acceleration
for $z<0.5 $ and some preliminary evidence of deceleration for
$z>0.5$ \cite{turner01}. If we assume that $\rho _{\varphi }\simeq
\rho _{m}$ at $z=0.5$, we can explain the current acceleration of
the universe using the same field that drove inflation. Because at
small redshift the curvature could be relevant, we found for
$z<0.5$ the field $\varphi $ makes the universe look flat,
although closed. The evolution through the matter domination to
the present is very similar to what is shown in Figure 1.

\section{Discussion}

I have developed a consistent observational model based on the
Holographic Principle which explains the role of inflation. It uses
the HP as a relevant principle for inflation and leads to a scenario
where it is possible to explain the current observation of
accelerated expansion. In this way it reduces the amount of scalar
fields needed to explain inflation and the dark energy. If reheating
is efficient enough to change the relative weight of energy
densities, it can also explain the coincidence problem. I have not
used a particular potential form, using instead the inflaton
equation of state as the relevant object of study. After reheating,
the universe becomes radiation-dominated at $z\sim 20 $, and the
inflaton energy density remains constant during all this period,
making the model consistent with nucleosynthesis and making it
behave similarly to the $\Lambda$CDM model. A better mechanism to
stopped the inflaton decay is a matter of current research
\cite{relicfi4,relicfi5}, that can be adopted by our model. The main
idea to get the inflation relic domination after $z \sim 0.5 $, and
its main quality -- that of connect inflation and the dark energy
with holographic properties of the universe -- are of sufficient
interest to deserve further study.

\section*{Acknowledgments}

The author wishes to thank S. del Campo for valuable discussions.

\end{document}